\pdfoutput=1
\documentclass[epsfig,12pt]{article}
\usepackage{epsfig}
\usepackage{graphicx}

\usepackage{latexsym}
\usepackage{amsmath}
\usepackage{amssymb}
\usepackage{relsize}
\usepackage{geometry}
\usepackage{color}
\def\beq{\begin{equation}}
\def\eeq{\end{equation}}
\def\beqn{\begin{eqnarray}}
\def\eeqn{\end{eqnarray}}

\newcommand{\ntwo}{${\mathcal N}=2\,$}

\newcommand{\gsim}{\lower.7ex\hbox{$
\;\stackrel{\textstyle>}{\sim}\;$}}
\newcommand{\lsim}{\lower.7ex\hbox{$
\;\stackrel{\textstyle<}{\sim}\;$}}

\def\beqn{\begin{eqnarray}}
\def\eeqn{\end{eqnarray}}

\def\beq{\begin{equation}}
\def\eeq{\end{equation}}
\def\ba{\beq\new\begin{array}{c}}
\def\ea{\end{array}\eeq}

\def\slashed#1{\setbox0=\hbox{$#1$}             % set a box for #1
   \dimen0=\wd0                                 % and get its size
   \setbox1=\hbox{/} \dimen1=\wd1               % get size of /
   \ifdim\dimen0>\dimen1                        % #1 is bigger
      \rlap{\hbox to \dimen0{\hfil/\hfil}}      % so center / in box
      #1                                        % and print #1
   \else                                        % / is bigger
      \rlap{\hbox to \dimen1{\hfil$#1$\hfil}}   % so center #1
      /                                         % and print /
   \fi}                                        %

\begin{document}

%%%%%%%%%%%%%%%%%%%%%%%%%%%%%%%%

\begin{titlepage}

\begin{flushright}
FTPI-MINN-10/17, 
UMN-TH-2909/10\\
July 2
\end{flushright}

\vspace{1cm}

\begin{center}
{  \Large \bf  Understanding Confinement in QCD:\\[3mm]
    Elements of a Big Picture}
\end{center}

\vspace{1mm}

\begin{center}

 {\large
 \bf    Mikhail Shifman}
\end {center}

\begin{center}

%\vspace{3mm}

{\it  William I. Fine Theoretical Physics Institute,
University of Minnesota,
Minneapolis, MN 55455, USA}

\end{center}

\vspace{5cm}

\begin{center}
{\large\bf Abstract}
\end{center}

I give a brief review of advances in the strong interaction theory.
This 
talk was delivered at the Conference in honor of
Murray Gell-Mann's 80th birthday,
 24--26 February 2010, Singapore. 

\end{titlepage}

\section{Introduction}
\label{intro}

\begin{figure}
\epsfxsize=4.5cm
%\centerline{\epsfbox{nsigma22.eps}}
\centerline{\epsfbox{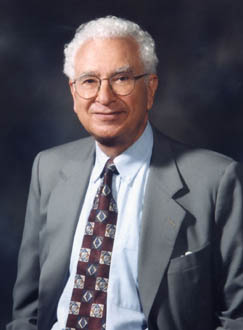}}
\caption{\small  Everybody knows that this is Murray Gell-Mann.}
\label{mgm}
\end{figure}

 In the early 1970s, when 
quantum chromodynamics (QCD) was born,
I was just in the beginning of my career in theoretical physics. My teachers at 
ITEP\,\footnote{Institute for Theoretical and Experimental Physics in Moscow}
 tried to convey to me a number of ``commandments" which were intended for guidance in my future
 scientific life. One of them was: always listen to what Gell-Mann says because he has a direct line to God.
 I always did. Gell-Mann was one of the discoverers of QCD who opened a whole new world.
 Unlike many recent theoretical constructions, whose relevance to nature is still a big question 
 mark, QCD will stay with us forever. I am happy that I invested so much time and effort in studying QCD.
 This was a long and exciting journey.
Almost 40 years later, I am honored and proud to be invited to 
this Conference
celebrating Professor Gell-Mann's 80$^{\rm th}$ birthday to give a talk on  advances in QCD.

 \begin{figure}[h]
\epsfxsize=15.5cm
%\centerline{\epsfbox{nsigma22.eps}}
\centerline{\epsfbox{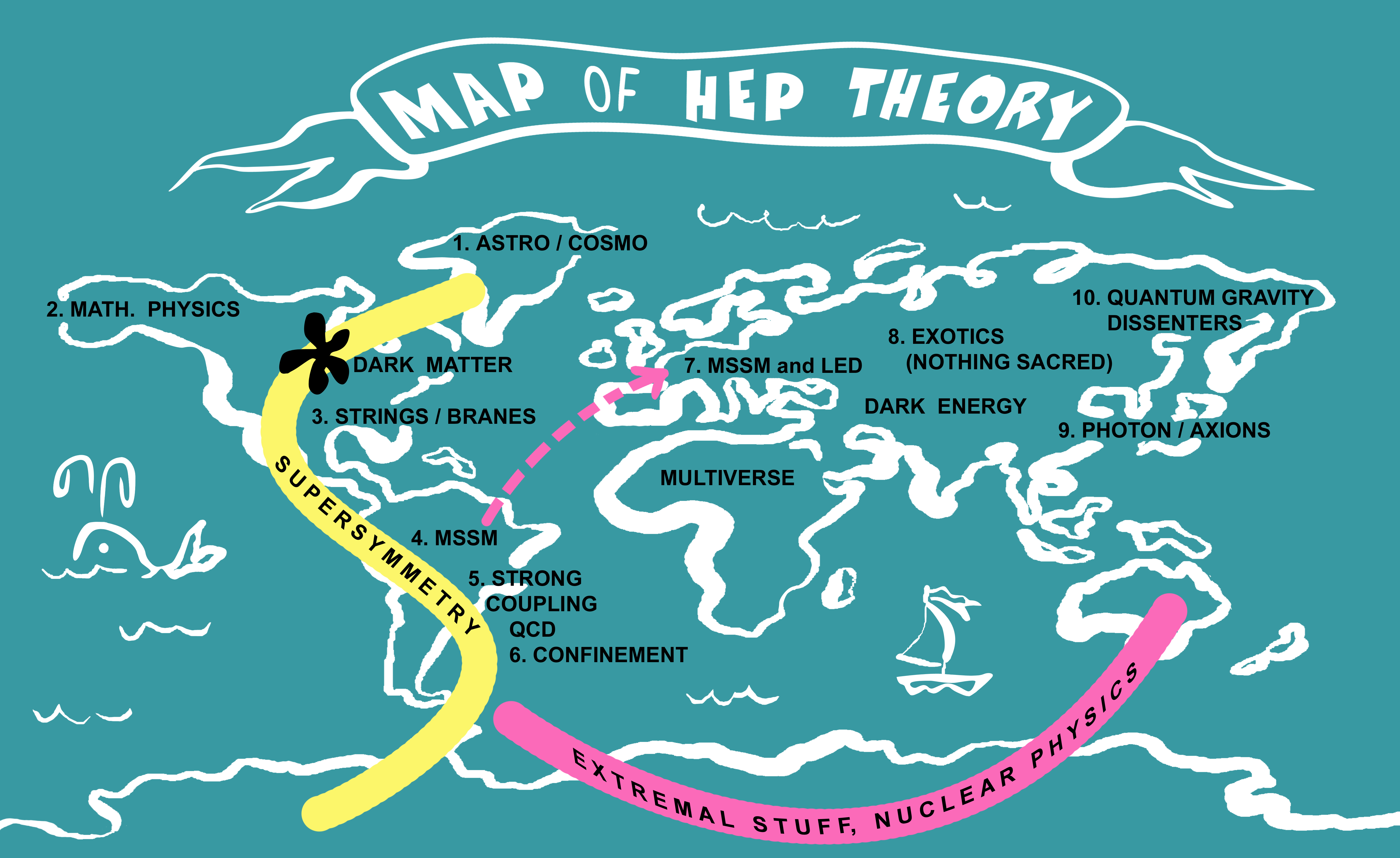}}
\caption{\small Map of the HEP theory. }
\label{mapp}
\end{figure}

I should say that the problem of strong interactions turned out to be extremely difficult (despite the fact that the underlying Lagrangian is firmly established) and 
the advances slow and painful. This is a usual story with the strong  coupling regime:
whenever theorists find themselves at strong coupling, they are in trouble. Yang--Mills theories
are no exception.

My task today is to outline some contours of the strong interaction
theory which gradually emerged from obscurity
during these four decades. Yes, the theory is incomplete, but those parts which are already in existence
are beautiful, and continue to grow.

First, to give a general idea of the role 
 which this theory plays in high energy physics (HEP)
I would like to chart approximate contours 
of main areas that
are under intense development in the theoretical community of today.
To this end  I display a symbolic map in Fig. \ref{mapp}. The linkage of the HEP theory map to 
the earth's geography is arbitrary and does not mean anything. You should pay attention 
only to interconnections of various areas of HEP. You see that QCD and strongly coupled
gauge theories at large, in which Murray Gell-Mann was a  trailblazer,
%\footnote{In fact, he was one of the discoverers of the whole continent.} 
occupy a vast area.
The advances there and in neighboring areas crucially depend on the exchange of ideas between them.
One can say that they feed each other. Of particular importance was a breakthrough impact of
supersymmetry of which I will speak later. In the Appendix you can find some additional information regarding the HEP theory map obtained from a HEP world traveller. Unfortunately, the traveller
was unable to visit some areas (allegedly, because of clearance issues), namely, that
of extremal phenomena (high-energy and high density QCD), nuclear physics, multiverses,
and theoretical nonperturbative supersymmtry.

\section{QCD}
\label{qcdd}
 
 In all processes with hadron participation strong interactions play a role.
 All matter surrounding us is made  of protons and neutrons лл the most common representatives
 of the class of hadrons.
 Even if you do not see them explicitly, they will necessarily show up at a certain stage or in loops.
 The fundamental Lagrangian governing strong interactions is
 \beq
 {\mathcal L} = \sum_f \bar{\psi}_f\left( i \,/\!\!\!\! D - m_f\right)\psi^f 
 -\frac 1 4 G_{\mu\nu}^a \,G^{\mu\nu\, a}\,.
 \label{govl}
 \eeq
 The first term describes color-triplet quarks and their coupling to color-octet
 gluons. It ascends to Gell-Mann. The second Yang--Mills term describes the gluon dynamics.
 Both terms taken together comprise the Lagrangian of quantum chromodynamics (QCD).
 The very name ``quantum chromodynamics" ascends to Gell-Mann too.
 Much in the same way as the Schr\"{o}dinger equation codes all of quantum chemistry,
 the QCD Lagrangian codes all of 
 
$\bullet$  nuclear physics;

$\bullet$ Regge behavior;

$\bullet$ neutron stars;

$\bullet$ chiral physics;

$\bullet$ light \& heavy quarkonia;

$\bullet$ glueballs \& exotics;

$\bullet$ exclusive \& inclusive hadronic scattering at large momentum transfer;

$\bullet$ interplay between strong forces \& weak interactions,

$\bullet$ quark-gluon plasma;

and much more. 

Although the underlying Lagrangian (\ref{govl}),  and asymptotic freedom it implies at short
distances \cite{AF},  are established beyond any doubt, the road from this 
starting point to theoretical control over the 
large-distance hadronic world is long and difficult. The journey which started 40 years
ago is  not yet completed. {\em En route}, many beautiful theoretical constructions
were developed allowing one to understand various corners of the hadronic world.
Here I am unable even to list them, let alone discuss in a comprehensible way. Therefore, I will focus
only on one -- albeit absolutely global  -- aspect defining the hadronic world: the confinement phenomenon.

 \section{Confinement in Non-Abelian Gauge Theories: dual Meissner Effect}
\label{cinagt}

%\zerocounters

The most salient feature of pure Yang-Mills theory  is linear confinement.
 If one takes a heavy probe quark and
an antiquark separated by a large distance, the force between them does not fall off with distance,
while the potential energy grows linearly. This is the explanation of the empiric fact
that quarks and gluons (the microscopic degrees of freedom in QCD)
never appear as asymptotic states. The physically observed spectrum consists of color-singlet
mesons and baryons. The phenomenon got the name color confinement, or, in a more narrow sense,
quark confinement. In the early days of QCD it was also referred to as
infrared slavery.

Quantum chromodynamics (QCD), and Yang--Mills theories at strong coupling at large, are not yet
analytically solved. Therefore, it is reasonable to ask:

Are there physical phenomena in which interaction energy between two interacting bodies grows with distance at large distances? Do we understand the underlying mechanism?

\begin{figure}[h]
\epsfxsize=11cm
%\centerline{\epsfbox{nsigma22.eps}}
\centerline{\epsfbox{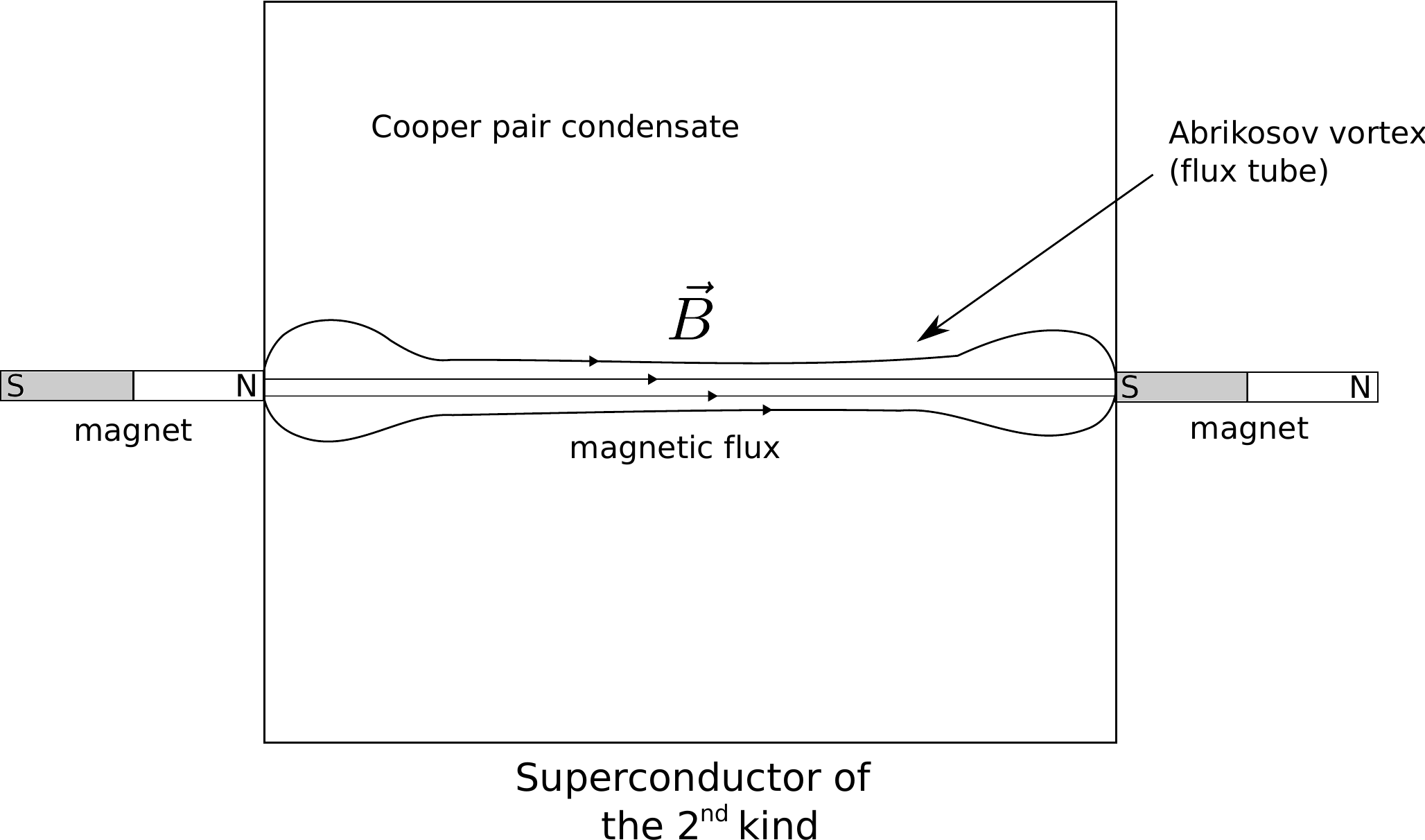}}
\caption{\small The Meissner effect in QED. }
\label{maisne}
\end{figure}

The answer to these questions is positive. The phenomenon of linearly growing potential
 was predicted by Abrikosov 
\cite{ANO} in the superconductors of the second type. The corresponding set up is shown 
in Fig.~\ref{maisne}.
In the middle of this figure we see a superconducting sample, with two very long magnets attached to it.
The superconducting medium does not tolerate the magnetic field. On the other hand, the flux of the magnetic field
must be conserved. Therefore, the magnetic field lines 
emanating from the $N$ pole of one magnet
find their way to the $S$ pole of another magnet, through 
the medium, by virtue of a flux tube formation.
Inside the flux tube the Cooper pair condensate vanishes 
and superconductivity 
is destroyed. The flux tube has a fixed tension, implying a constant force between the
magnetic poles as long as they are inside the superconducting sample. 
The phenomenon described above is sometimes referred to as the Meissner effect.

Of course, the Meissner effect of the Abrikosov  type
occurs in the Abelian theory, QED. The flux tube that forms in this case is Abelian.
In Yang--Mills theories we are interested in   non-Abelian analogs of the Abrikosov vortices.
Moreover, while in the Abrikosov case the flux tube is that of the magnetic field,
in QCD and QCD-like theories the confined objects are quarks;
therefore, the flux tubes must be ``chromoelectric" rather than chromomagnetic.
In the mid-1970s Nambu,   't Hooft, and  Mandelstam (independently) 
put forward an idea \cite{NTM}
of a ``dual Meissner effect" as the underlying mechanism for color confinement.
Within their conjecture, in chromoelectric theories ``monopoles" condense
leading to formation of ``non-Abelian flux tubes" between the probe quarks.
At this time the  Nambu--'t Hooft--Mandelstam paradigm was not even a physical 
scenario, rather a
dream, since people had no clue as to
the main building blocks such as non-Abelian  
flux tubes. After the Nambu--'t Hooft--Mandelstam
conjecture had been formulated many works were published on this subject, to no avail.

A long-awaited breakthrough discovery came 20 years later: 
 the Seiberg--Witten solution \cite{SW} of
${\mathcal N}=2$ super-Yang--Mills theory
 slightly deformed by a superpotential breaking ${\mathcal N}=2$ down to ${\mathcal N}=1$.
In the ${\mathcal N}=2$ limit, the theory has a moduli space. If the gauge group is 
 SU(2),
on the moduli space,
SU(2)$_{\rm gauge}$ is spontaneously broken down to U(1). Therefore, the theory possesses 
the 't Hooft--Polyakov monopoles \cite{TP} in the quasiclassical regime.
Of course, in  this regime they are very heavy and play no role in dynamics.
 Using the power of ${\mathcal N}=2$
supersymmetry, two
special  points on the moduli space were found \cite{SW} at strong coupling,
(the monopole and dyon points),
in which the monopoles (dyons) become massless. 
In these points  the scale of the gauge symmetry breaking 
\beq
{\rm SU}(2)\to {\rm U}(1)
\label{one}
\eeq
 is determined by $\Lambda$, the dynamical scale parameter
 of ${\mathcal N}=2$ super-Yang--Mills theory. 

All physical states can be classified with regards to the unbroken
U(1). It is natural to refer to the U(1) gauge boson as to the photon. In addition to the photon, all
its superpartners, being  neutral,  remain  massless at this stage, while all other states, with nonvanishing 
``electric" charges,
 acquire masses of the order of $\Lambda$.
 In particular, two gauge bosons corresponding to SU(2)/U(1) --
 it is natural to call them $W^\pm$ -- have masses $\sim \Lambda $.
All such states  are ``heavy" and can be integrated out.
 
In the low-energy limit, near the monopole and dyon points,
one deals with electrodynamics of massless monopoles. One can formulate an effective local
theory describing interaction of the light states by dualizing the original
phton.  This is a U(1) gauge  theory 
in which the (magnetically) charged matter fields $M$, $\tilde M$ are those of monopoles while the
U(1) gauge field that couples to $M$, $\tilde M$
 is {\em dual} with respect to the photon of the original theory.
 The ${\mathcal N}=2$
preserving  superpotential has the form ${\mathcal W} = {\mathcal A} M\tilde M$,
where ${\mathcal A}$ is  the ${\mathcal N}=2$ superpartner of the dual photon/photino fields.

Now, if one switches on a small ${\mathcal N}=2$ breaking superpotential,
the only change in the low-energy theory is the emergence of the extra $m^2{\mathcal A} $
term in the superpotential. Its impact is crucial: it triggers the monopole condensation,
$\langle M\rangle = \langle \tilde M\rangle =m$, which implies, in turn,
that the dual U(1) symmetry is spontaneously broken, and the dual photon acquires a mass
$\sim m$. As a consequence, the Abrikosov flux tubes
are formed. Viewed inside the dual theory, they carry fluxes
of the magnetic field. With regards to the original microscopic theory
these are the electric field fluxes.

Thus, Seiberg and Witten demonstrated, for the first time ever, the existence of the dual Meissner
effect in a judiciously chosen non-Abelian gauge field theory. 
If one ``injects" a probe (very heavy) quark and antiquark in this theory,
a flux tube  forms
between them, with necessity,
leading to linear confinement.

The flux tubes in the Seiberg--Witten solution were investigated in detail in
\cite{HSZ}.
These flux tubes are Abelian, and so is confinement caused by their formation.
What does that mean? At the scale of distances at which the flux tube is formed
(the inverse mass of the Higgsed U(1) photon) the gauge group that is operative
is Abelian. In the Seiberg--Witten analysis this is the dual U(1). 
The off-diagonal (charged) gauge bosons are very heavy in this scale
and play no direct role in the flux tube formation and 
confinement that ensues. Naturally, the spectrum of composite objects in this case
turns out to be richer than that in QCD and similar theories with non-Abelian
confinement. By  non-Abelian
confinement I mean such dynamical regime in which at distances
of the flux tube formation all gauge bosons are equally important.

 Moreover, the string topological stability is based
on $\pi_1 ({\rm U}(1)) = {\mathbb Z}$. Therefore, $N$ strings  
do not annihilate as they should in SU$(N)$ QCD-like theories. 

The two-stage symmetry breaking pattern, with ${\rm SU}(2)\rightarrow {\rm U}(1)$ 
occurring at a high scale
while the dual U(1) $\to$ nothing 
 at a much lower scale, 
 has no place in QCD-like theories, as we know  from experiment. In such theories, presumably, 
 all non-Abelian gauge degrees of freedom
take part in the string formation,
and are operative at the scale at which the strings are formed. 
The strings  in the 
Seiberg--Witten solution are believed to
belong to the same universality class as those in  QCD-like theories.
However, in the limit of large-$m$ deformations, when a non-Abelian regime presumably sets
in and non-Abelian strings develop in the model considered by Seiberg and Witten, theoretical control is completely lost. Thus, 
the status of the statement of the same universality class  is conjectural.

\section{Non-Abelian strings}
\label{nastr}

In a bid to better understand string-based confinement mechanism
in Yang--Mills theories that might be more closely related to QCD
people continued searches for models supporting non-Abelian strings. If
a model in which non-Abelian strings develop in a fully controllable manner, i.e. at weak coupling,
could be found and the passage from Abelian to non-Abelian strings explored, this would provide
us with evidence that no phase transition occurs between the two regimes in the Seiberg--Witten 
solution.

\begin{figure}[h]
\epsfxsize=8cm
%\centerline{\epsfbox{nsigma22.eps}}
\centerline{\epsfbox{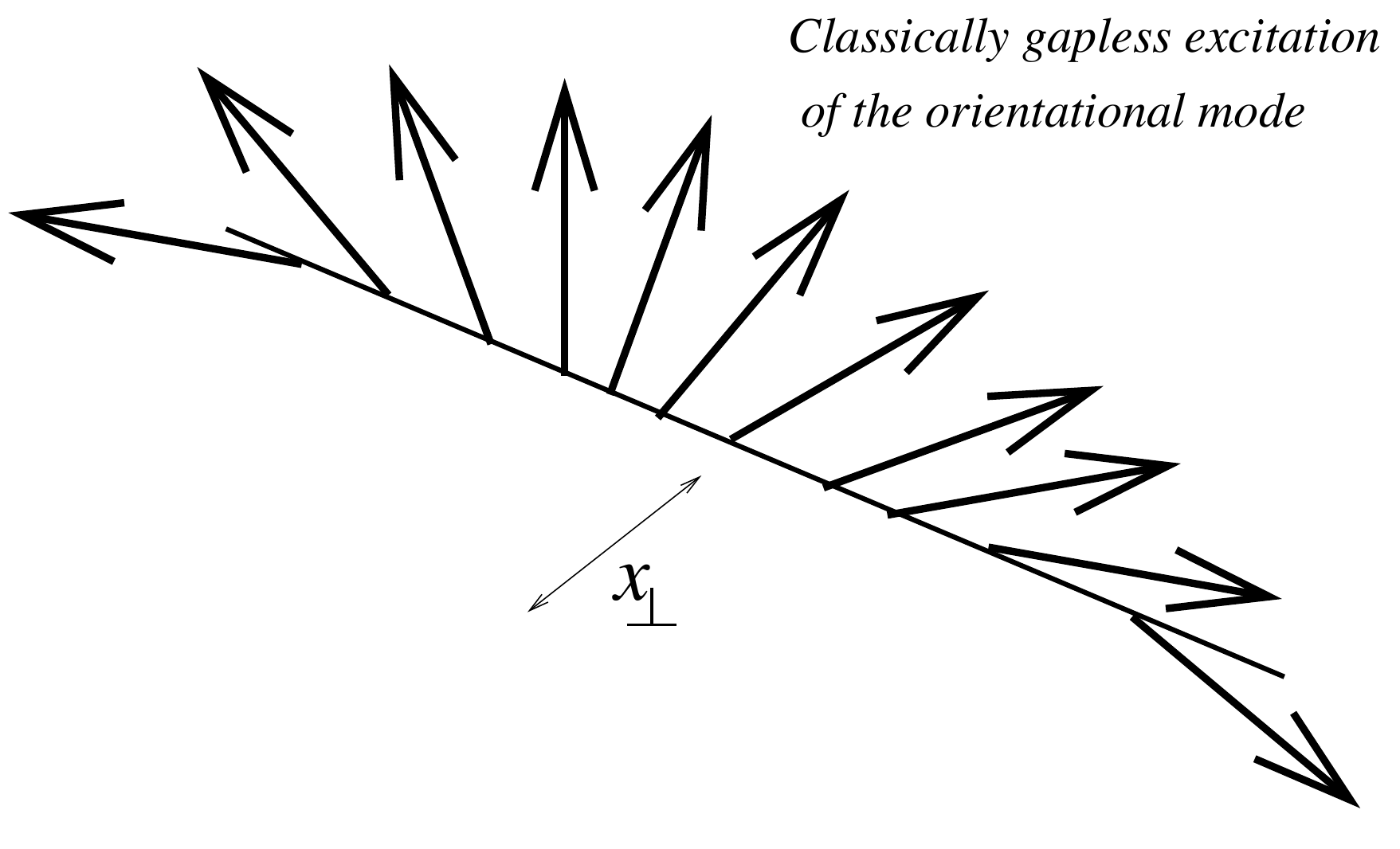}}
\caption{\small Orientational moduli on the string world sheet. }
\label{maisnep}
\end{figure}

In the technical sense,  what does one mean when one speaks of non-Abelian flux tubes?
Apparently, the orientation of the magnetic field in the tube  interior must be free
to strongly fluctuate inside the SU$(N)$ group. There is no such freedom in the Abelian string
of the Abrikosov type. In other words, in addition to translational moduli,
the theory on the string world sheet must acquire orientational moduli (Fig.~\ref{maisnep}).

If one thinks there is a kind of string theory behind QCD confining dynamics, such behavior is natural. Indeed, string theory is formulated in higher dimensions. Bringing it to $D=4$ requires compactification of some dimensions. If (some of) compact dimensions have isometries, the corresponding sigma model on the 4D string world sheet will have classically massless internal degrees of freedom. For instance, 
compactification on $S_2 $ gives rise to  ${\rm  CP}(1)$ sigma model
on the string world sheet. Two-dimensional infrared dynamics will then generate a mass gap for these orientational degrees of freedom.

That's why searches for non-Abelian
flux tubes and non-Abelian mono\-poles in the bulk Yang--Mills theories continued,
with a decisive breakthrough in 2003-04 \cite{HT,auzzietal}. By that time the program of
finding field-theoretical analogs of all basic constructions
of string/D-brane theory was in full swing.
BPS domain walls, analogs of D branes, had been identified in
supersymmetric Yang--Mills theory \cite{DS}. It had been demonstrated that
such walls support gauge fields localized on them. BPS saturated 
string-wall junctions had been constructed \cite{SY1}. Topological stability of the 
non-Abelian strings under consideration is due to the fact that
\beq
\pi_1\left( \frac{\rm {SU}(2) \times \rm{U}(1)}{Z_2} \right)  \,\to \, \mbox{nontrivial}\,.
\eeq

\section{Basic Bulk Theory: setting the stage}
\label{naft}

Non-Abelian strings were first found 
in ${\mathcal N}=2$  super-Yang--Mills theories with U(2)$_{\rm gauge}$ and two matter hypermultiplets \cite{HT,auzzietal}. 
The \ntwo vector multiplet
consists of the  U(1)
gauge field $A_{\mu}$ and the SU(2)  gauge field $A^a_{\mu}$,
(here $a=1,2,3$), and their Weyl fermion superpartners
($\lambda^{1}$, $\lambda^{2}$) and
($\lambda^{1a}$, $\lambda^{2a}$), plus
complex scalar fields $a$, and $a^a$.  The  global SU(2)$_R$ symmetry inherent to
 \ntwo   models manifests itself through rotations
$\lambda^1 \leftrightarrow \lambda^2$.

The quark multiplets consist
of   the complex scalar fields
$q^{kA}$ and $\tilde{q}_{Ak}$ (squarks) and
the  Weyl fermions $\psi^{kA}$ and
$\tilde{\psi}_{Ak}$, all in the fundamental representation of 
the SU(2) gauge group
($k=1,2$ is the color index
while $A$ is the flavor index, $A=1,2$).
The scalars $q^{kA}$ and ${\bar{\tilde q}}^{\, kA}$
form a doublet under the action of the global
SU(2)$_R$ group.  The quarks and squarks have a U(1) charge too.

\begin{figure}
\begin{center}
\psfig{figure=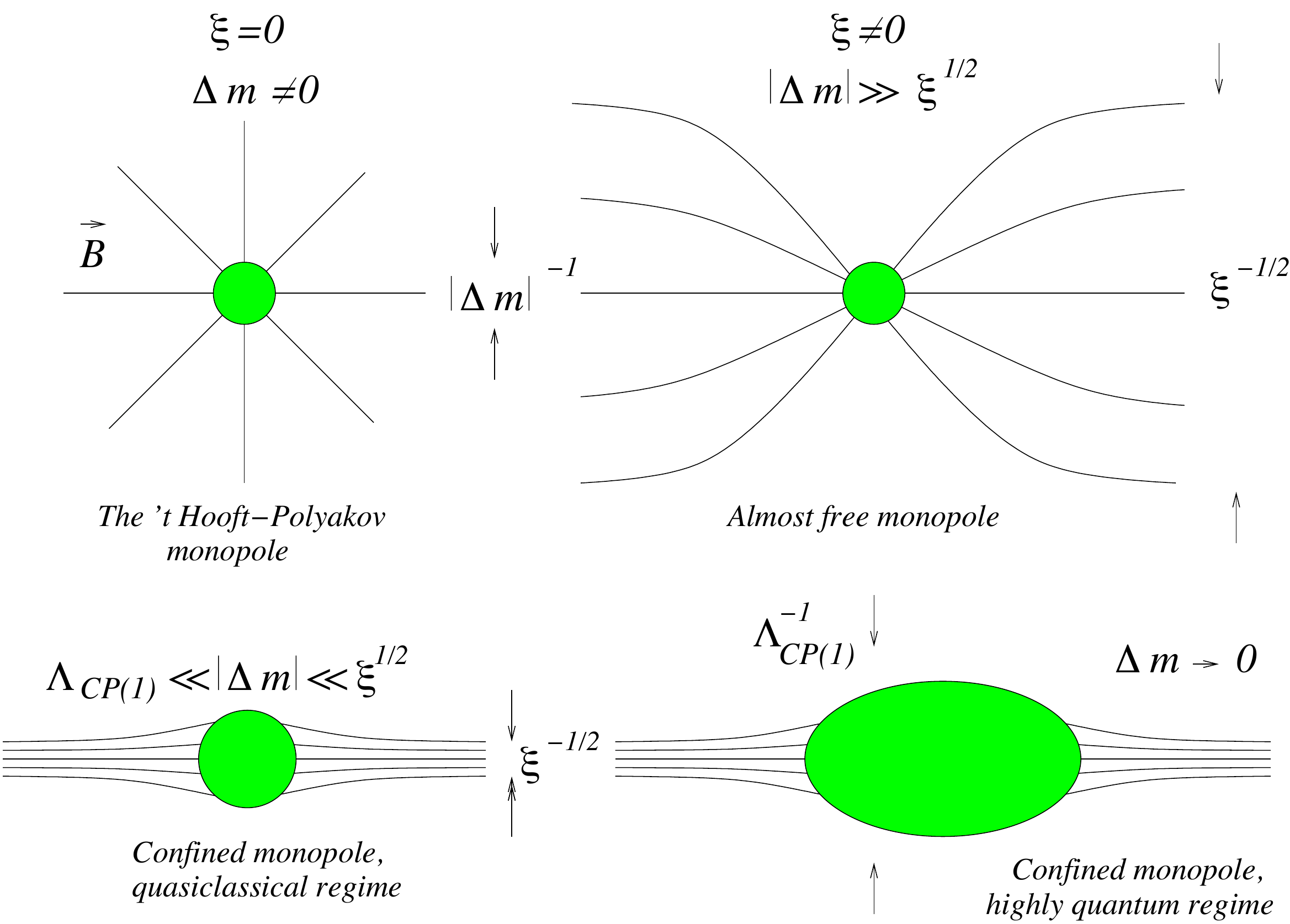,height=2.5in}
\end{center}
\caption{\small Various regimes for monopoles and strings.}
\label{fig:radish}
\end{figure}

If one introduces a non-vanishing Fayet--Iliopoulos parameter $\xi$
the theory develops isolated quark vacua,
in which the gauge symmetry is fully Higgsed, and all elementary excitations are massive.
In the general case, two matter
mass terms allowed by  ${\mathcal N}=2$
are unequal, $m_1\neq m_2$. 
There are free parameters whose interplay
determines dynamics of the theory:
the Fayet--Iliopoulos parameter $\xi$, the mass difference
$\Delta m$ and a dynamical scale parameter
$\Lambda$, an analog of the QCD scale $\Lambda_{\rm QCD}$ (Fig.~\ref{fig:radish}). 
Extended supersymmetry guarantees that some crucial dependences are holomorphic,
and there is no phase transition.

Both the gauge and flavor symmetries of the model are
broken by the squark condensation. 
All gauge bosons acquire the same masses (which are of the order of 
inverse string thickness). A global diagonal
combination of color and flavor groups, SU$(2)_{C+F}$, survives the breaking
(the subscript $C+F$ means a combination of global color and flavor groups). 

While SU$(2)_{C+F}$ is
the symmetry of the vacuum, the flux tube solutions break it spontaneously.
This gives rise to orientational moduli on the string world sheet. 

The bulk theory is characterized by three parameters of dimension of mass:
$\xi$, $\Delta m$, and $\Lambda$.
As various parameters vary, the theory under consideration evolves in a very
graphic way, see Fig.~\ref{fig:radish}. At $\xi=0$ but 
$\Delta m \neq 0$
(and $\Delta m \gg \Lambda$) it presents a very clear-cut example
of a model with the standard 't Hooft--Polyakov monopole.
This is due to the fact that the relevant part of the bosonic sector is nothing but the
Georgi--Glashow model.
The monopole is unconfined  --- the flux tubes are not yet formed.

Switching on $\xi\neq 0$ traps the magnetic fields inside
the flux tubes, which are weak as long as $\xi\ll\Delta m$.
The flux tubes change the shape of the monopole far away from its core,
leaving the core essentially intact. Orientation of the
chromomagnetic field inside the flux tube is essentially fixed. This is due to the fact that all off-diagonal gauge bosons ($W$ bosons) are heavy in this limit.
Thus, the flux tubes supported in this limit are {\em Abelian}. (They are commonly referred to as the 
$Z_N$ strings.)

With $|\Delta m|$ decreasing, 
fluctuations in the orientation of the 
chromomagnetic field inside the flux tubes grow. 
Simultaneously, the monopole 
which no loner resembles the 't Hooft--Polyakov monopole, is 
seen as a {\em string junction}.

Finally, in the limit $\Delta m\to 0$
the transformation is complete. A global SU(2) symmetry restores
in the bulk. All three gauge bosons have identical masses.
Orientational (exact, classically massless) moduli
develop on the string world sheet making it non-Abelian.
The string world sheet theory is CP(1) (CP$(N-1)$ for 
generic values of $N$). 
Two-dimensional CP$(N-1)$ models with four supercharges
are asymptotically free. They have $N$ distinct vacuum states.

Each vacuum state of the worldsheet CP$(N-1)$ theory
presents a distinct string from the standpoint of the bulk theory.
There are $N$ species of such strings; they have degenerate 
tensions $T_{\rm st} =2\pi\xi$.
The ANO string tension is $N$ times larger.

\begin{figure}
\begin{center}
\psfig{figure=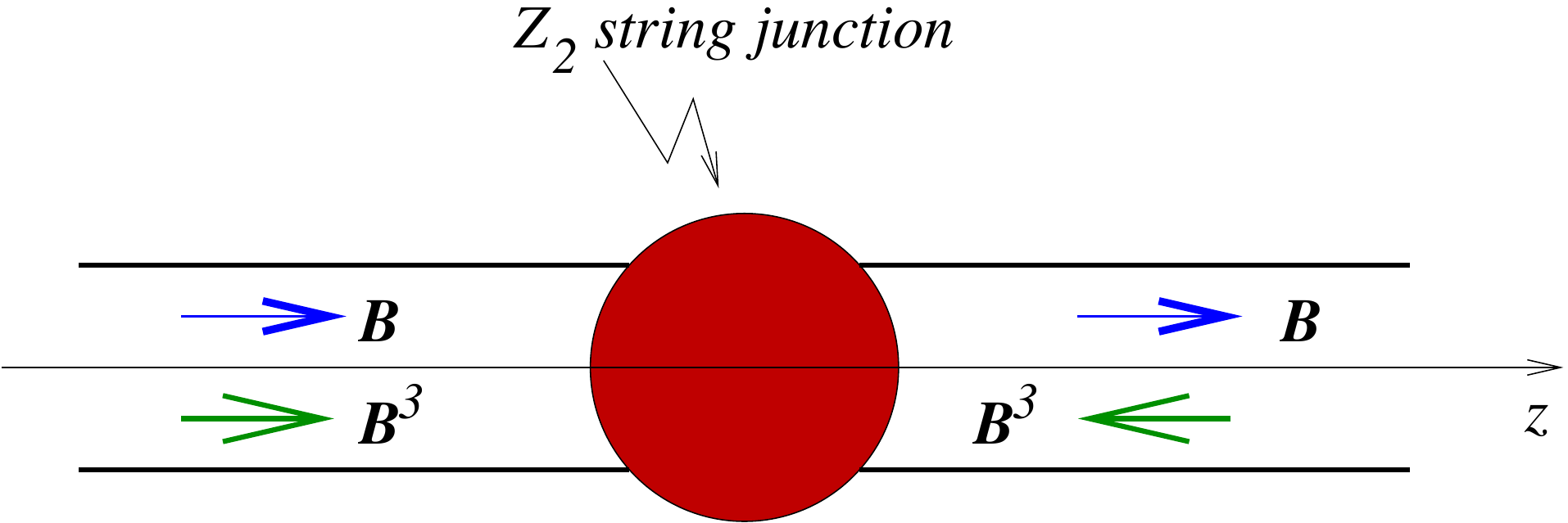,height=1in}
\end{center}
\caption{\small $Z_2$  string junction.}
\label{z2sj}
\end{figure}

Two different strings can form a stable junction.
Figure~\ref{z2sj} shows this junction in the limit
\beq
\Lambda_{{\rm CP}(1)}\ll |\Delta m| \ll \sqrt{\xi}
\label{limi}
\eeq
corresponding to the lower left corner in Fig.~\ref{fig:radish}.
The magnetic fluxes of the U(1) and SU(2) gauge groups
are oriented along the $z$ axis. In the limit (\ref{limi})
the SU(2) flux is oriented along the third axis in the internal space.
However, as $|\Delta m| $ decreases, fluctuations
of $B_z^a$  in the internal space grow, and at $\Delta m\to 0$
it has no particular orientation in SU(2) (the lower right corner of Fig.~\ref{fig:radish}).
In the language of the world-sheet theory this phenomenon
is due to restoration of the O(3) symmetry in the quantum vacuum of
the CP(1) model. 

Evolution from the upper right corner in Fig.~\ref{fig:radish} to the lower right corner
is in fact the transformation of the Abelian string into non-Abelian. 
${\mathcal N}=2$ supersymmetry guarantees that it is smooth, with no phase transition.

The junctions of degenerate strings present what remains of the
monopoles in this highly quantum regime \cite{SY2,HT2}. It is remarkable that,
despite the fact we are deep inside the highly quantum regime,
holomorphy allows one to exactly calculate the mass of
these monopoles. This mass is given by the expectation value of the
kink central charge in the worldsheet CP$(N-1)$ model
(including the anomaly term), $M_M \sim N^{-1}\, \langle R \,  \psi_L^\dagger\,\psi_R\rangle$. 
 
 \section{Towards ${\mathcal N}=1$}
 \label{towar}
 
The ``unwanted" feature of ${\mathcal N}=2$ theory,
making it less similar to QCD than one would desire, is the presence of the adjoint chiral
superfields ${\mathcal A}$ and ${\mathcal A}^a$. One can get rid of them making them heavy. 
To this end we can endow the adjoint superfield
with a mass term of the type
$\mu {\mathcal A}^2$, through the ${\mathcal N}=1$
preserving superpotential
\beq
{\mathcal W} = \frac{\mu}{2}\left[ {\mathcal A}^2 + \left({\mathcal A}^a\right)^2\right]\,.
\label{mcsp}
\eeq
Now, supersymmetry of the bulk model becomes ${\mathcal N}=1$.
At large $\mu$ the adjoint fields decouple.

With the deformation  superpotential (\ref{mcsp})
the 1/2 BPS classical flux tube
solution stays the same as in the absence of this superpotential \cite{Shifman:2005st}. 
Moreover, the number of the boson and fermion zero modes,
which become moduli fields on the string world sheet,
does not change either. 
For the fermion zero modes this statement follows from an index theorem proved
in \cite{GSY}. If the string solution and the number of zero modes remain the
same, what can one say about the string world-sheet theory? 

The bulk deformation (\ref{mcsp}) leads to a  remarkable, {\em heterotic} deformation of the
CP(1) model on the world sheet, with ${\mathcal N}=(0,2)$ supersymmetry.
The discovery of non-Abelian strings in
${\mathcal N}=1$ bulk theories is a crucial step on the way to
the desired ${\mathcal N}=0$ theories. 
Moreover, the heterotically deformed CP(1) model is very rich by itself exhibiting
a number of distinct dynamical scenarios unknown previously.

To understand the emergence of ${\mathcal N}=(0,2)$ supersymmetry
in the world-sheet Lagrangian recall that
${\cal N}=2$  Yang--Mills theories which support
non-Abelian flux tubes have eight supercharges. 
The flux tube solutions are 1/2 BPS-saturated. Hence, 
the effective low-energy theory of the moduli fields
on the string world sheet must have four supercharges. 
The bosonic moduli consist of two groups: two translational moduli $\left(x_0\right)_{1,2}$
corresponding to translations in the plane perpendicular to the string axis,
and two orientational moduli whose interaction is described by CP(1) (see Fig.~\ref{maisnep}).
The fermion moduli also split in two groups:
four supertranslational moduli $\zeta_L ,\, \zeta_L^\dagger ,\,\zeta_R ,\, \zeta_R^\dagger$ 
plus four superorientational moduli.
${\cal N}=2$ supersymmetry in the bulk and on the worldsheet guarantees 
that $\left(x_0\right)_{1,2}$ and $\zeta_{L,R}$ form a free field theory on the worldsheet
completely
decoupling from (super)orientational moduli, which in turn form 
${\mathcal N}=(2,2)$ supersymmetric CP(1) model. 

When one deforms the bulk theory
to break ${\cal N}=2$ down to ${\cal N}=1$, one has four supercharges in the
bulk and expects two supercharges on the world sheet. 
Two out of four supertranslational modes, $\zeta_R $ and $\zeta_R^\dagger$,
get coupled to two superorientational modes $\psi_R $ and $\psi_R^\dagger$
\cite{ET}.
At the same time, $\zeta_L$ and $\zeta_L^\dagger$ remain protected.
Thus, the right- and left-moving fermions acquire different interactions; hence,
the flux tube becomes heterotic!

This breaks two out of four supercharges on the world sheet.
 Edalati and Tong outlined \cite{ET}
a general structure of the chiral ${\mathcal N}= (0,2)$ generalization
of CP(1). Derivation of the heterotic CP(1) model from the 
bulk theory was carried out in Ref.~\cite{shyu}.

\section{Heterotic non-Abelian string}
\label{hnas}

The Lagrangian of the heterotic CP$(N-1)$ model can be written as \cite{shyu}
\beqn
L_{{\rm heterotic}} && 
= 
\zeta_R^\dagger \, i\partial_L \, \zeta_R  + 
\left[\gamma\, g_0^2 \, \zeta_R  \, G_{i\bar j}\,  \big( i\,\partial_{L}\phi^{\dagger\,\bar j} \big)\psi_R^i
+{\rm H.c.}\right]
\nonumber
\\[4mm]
&&
 -g_0^4\, |\gamma |^2 \,\left(\zeta_R^\dagger\, \zeta_R
\right)\left(G_{i\bar j}\,  \psi_L^{\dagger\,\bar j}\psi_L^i\right)
\nonumber
\\[4mm]
&&
+G_{i\bar j} \big[\partial_\mu \phi^{\dagger\,\bar j}\, \partial_\mu\phi^{i}
+i\bar \psi^{\bar j} \gamma^{\mu} D_{\mu}\psi^{i}\big]
\nonumber
\\[4mm]
&&
- \frac{g_0^2}{2}\left( G_{i\bar j}\psi^{\dagger\, \bar j}_R\, \psi^{ i}_R\right)
\left( G_{k\bar m}\psi^{\dagger\, \bar m}_L\, \psi^{ k}_L\right)
\nonumber
\\[4mm]
&&
+\frac{g_0^2}{2}\left(1-2g^2_0|\gamma|^2\right)
\left( G_{i\bar j}\psi^{\dagger\, \bar j}_R\, \psi^{ i}_L\right)
\left( G_{k\bar m}\psi^{\dagger\, \bar m}_L\, \psi^{ k}_R\right).
\label{cpn-1g}
\eeqn

The constant $\gamma$ in Eq. (\ref{cpn-1g}) 
is the parameter which determines the ``strength" of the heterotic deformation,
and the left-right asymmetry in the fermion sector. It is related to
the parameter $\mu$ in Eq.~(\ref{mcsp}) (e.g. $\gamma\propto \mu$ at small $\mu$
).
The third, fourth and fifth  lines in Eq.~(\ref{cpn-1g}) are the same as in the
conventional ${\mathcal N}=(2,2)$ CP$(N-1)$ model, except the
last coefficient.

Introduction of a seemingly rather insignificant heterotic deformation
drastically changes dynamics of the CP(1) model, leading to spontaneous supersymmetry breaking. 
At small $\mu$ (small $\gamma$)
the field $\zeta_R$ represents a massless Goldstino, with the residue $\langle R \,  \psi_R^\dagger\,\psi_L\rangle$.  As well known,
a nonvanishing bifermion condensate $\langle R \,  \psi_R^\dagger\,\psi_L\rangle $ develops in the undeformed model. Thus, the vacuum energy 
\beq
{\mathcal E}_{\rm vac} = |\gamma|^2\,
\left|\langle R \,  \psi_R^\dagger\,\psi_L\rangle
\right|^2\neq 0\,.
\eeq
Therefore, upersymmetry is spontaneously broken.
 A nonvanishing ${\mathcal E}_{\rm vac}$ for arbitrary values of $\gamma$ in heterotically deformed 
 CP$(N-1)$ models was obtained in \cite{shyu2}
from the large-$N$ expansion.  
Spontaneous breaking of SUSY in heterotic $CP(N-1)$
was anticipated in \cite{Tong:2007qj}. 

\section{Large-$N$ solution of the heterotic CP$(N-1)$ model}
\label{largen}

\begin{figure}[h]
%\epsfxsize=6cm
%\centerline{\begin{rotate}{270}\epsfbox{phase_diag.eps}\end{rotate}}
\centerline{\includegraphics[width=8.5cm,keepaspectratio]{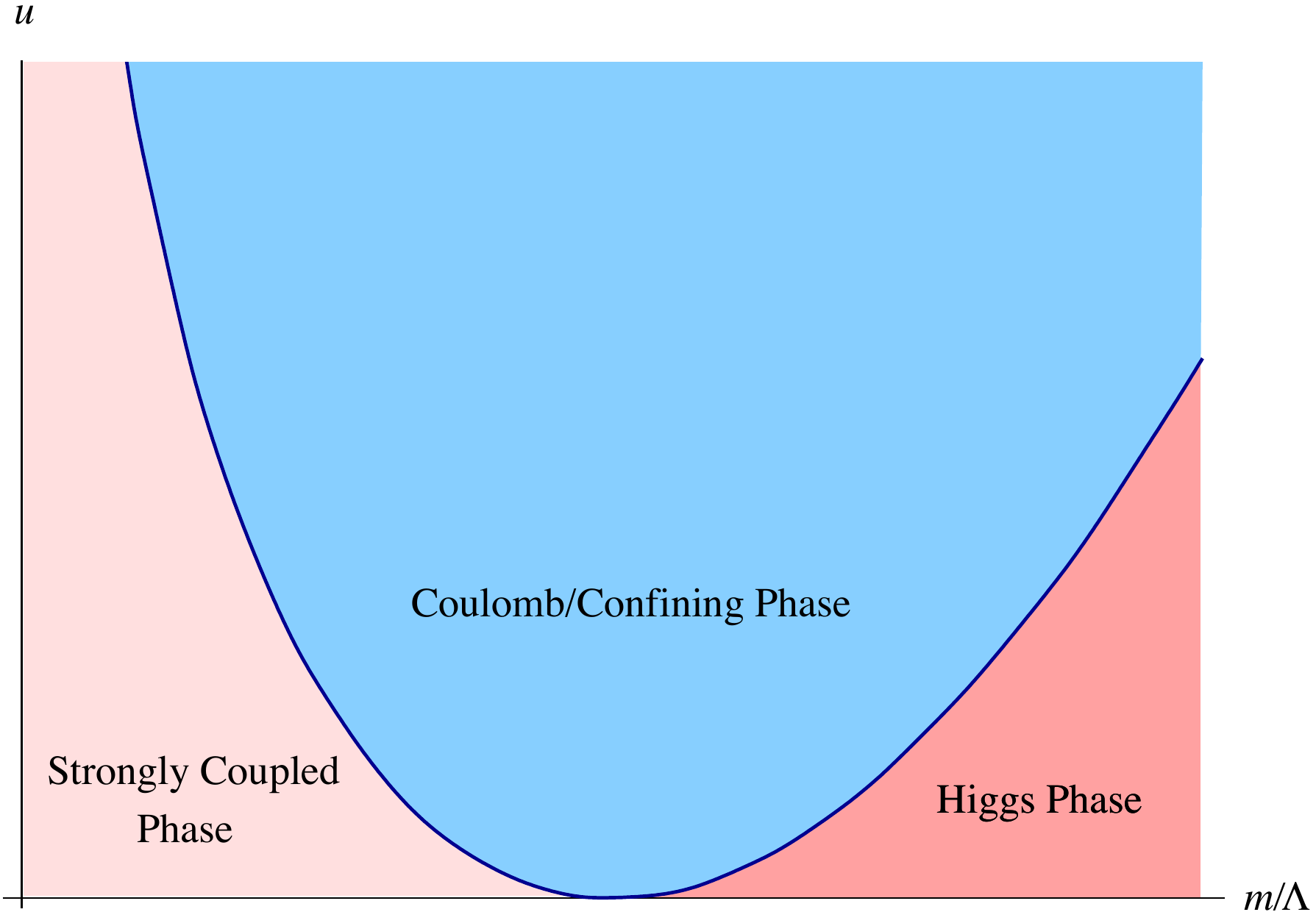}}
\caption{\small 
The phase diagram of the twisted-mass deformed heterotic CP($N-1$) theory.
The parameter $u$ denotes the amount of deformation and is related to $\gamma$.
}
\label{fig:higgsborder}
\end{figure}

To reveal a rich dynamical structure of the 
heterotically deformed CP$(N-1)$ models it is 
instructive to add twisted masses which correspond to $\Delta m\neq 0$ introduced above.
Moreover, the most convenient choice of the twisted masses is that
preserving the $Z_N$ symmetry of the model which exists at $\Delta m = 0$,
\beq
m_k = m \exp\left(i\frac{2\pi \,k}{N}
\right),\qquad k=0,1,2,..., N-1\,.
\label{jer1}
\eeq
where $m$ is a complex parameter setting the scale of the twisted masses. 
For simplicity I will take it real.
Now we have two variable
parameters, $m$ and $\gamma$, the strength of the heterotic deformation. The breaking vs. 
nonbreaking of the above $Z_N$ determines the phase diagram.
This model can be solved at large $N$ using the $1/N$ expansion \cite{shyu2}.
I will present here just two plots exhibiting main features of the solution.

\begin{figure}[h]
\epsfxsize=9cm
%\centerline{\epsfbox{nsigma22.eps}}
\centerline{\epsfbox{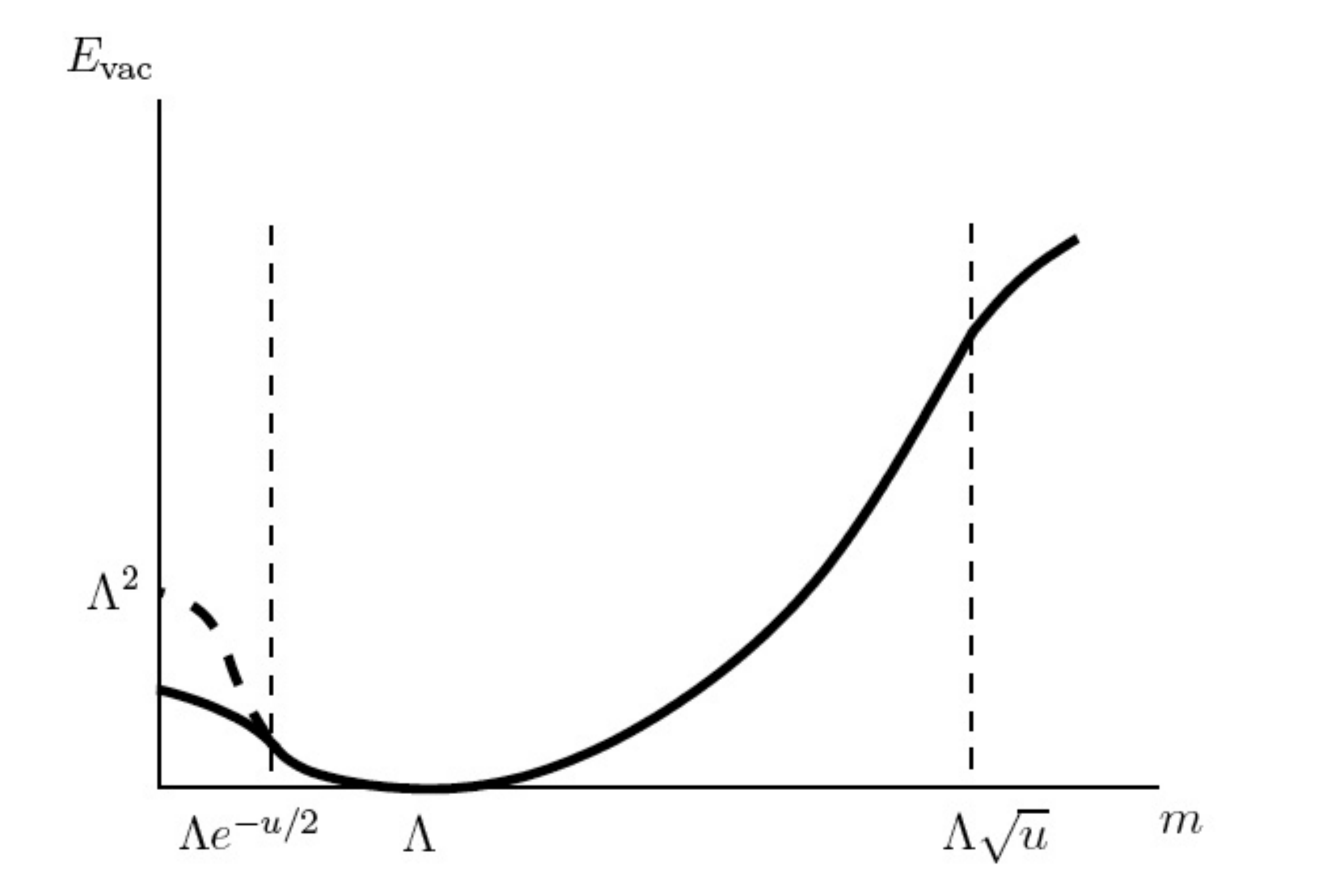}}
\caption{\small Vacuum energy density vs. $m$. The dashed line shows an unstable quasivacuum. }
\label{vachet}
\end{figure}

Figure~\ref{fig:higgsborder} displays three distinct regimes and two phase transition lines.
Two phases with the spontaneously broken $Z_N$ on the left and on the right
are separated by a phase with unbroken $Z_N$. This latter phase is characterized by a unique vacuum and confinement of all U(1) charged fields (``quarks").
In the broken phases (one of them is at strong coupling) there are $N$ degenerate vacua
and no confinement.

\begin{figure}[h]
\epsfxsize=4.5cm
%\centerline{\epsfbox{nsigma22.eps}}
\centerline{\epsfbox{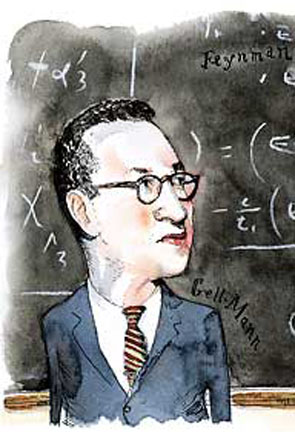}}
\caption{\small  Everybody knows that this is Murray Gell-Mann. (\copyright Barry Blitt, The Atlantic Monthly, July 2000.)}
\label{mgmat}
\end{figure}

Figure~\ref{vachet} shows the vacuum energy density at a fixed value of $\gamma$. It
demonstrates
that supersymmetry is spontaneously broken everywhere except a circle $|m|=\Lambda$ in the 
$Z_N$-unbroken phase. 
The first phase transition occurs
at strong  coupling (small $|m|$) while the second phase transition is at weak coupling
(large $|m|$).  
Both phase transitions between the three distinct phases  are of the second kind.

One must be able to translate this rich world sheet-dynamics
into (presumably) highly nontrivial statements regarding the bulk theory at strong coupling.

\section{Instead of conclusion}
\label{conc}

The progress in understanding dynamics of non-Abelian theories at strong coupling  was painfully slow. But what a progress it is! To properly appreciate the scale of achievements, please,    
 look back in the 1970's and compare what was known then about strong interactions 
to what we know now. Just open old reviews or textbooks devoted to this subject, in parallel with fresh publications. Of course, a pessimist might say that the full analytical theory is still elusive. 
Will it ever be created? And what does it mean, ``the full analytical theory," in the case when we are at strong coupling? The richness of the hadronic world is enormous.
Unlike QED we will never be able to analytically calculate
all physical observables with arbitrary precision. But do we really need this? 
To my mind, what is really needed is the completion
of the overall qualitative picture of confinement in non-supersymmetric theories,
supplemented by a variety of approximate quantitative tools custom-designed
to treat particular applications. A large number of such tools are already available.

\section*{Acknowledgments}

I am very  grateful to A. Yung with whom I shared the pleasure of working on the issues
discussed in this talk. Generous assistance of Andrey Feldshteyn
with cartoons is gratefully acknowledged.
This write-up was completed during my stay
at the Institut de Physique Th\'{e}orique, CEA-Saclay, which was made possible
due to the support of the 
Chaires Internacionales de Recherche Blaise Pascal, Fondation de l'Ecole Normale Sup\'{e}rieure.
This work is supported in part by the  DOE grant DE-FG02-94ER408.

\newpage

%\addcontentsline{toc}{section}{Appendices}

\section*{Appendix:  \!\! Snapshots made by  HEP world traveller}

 \renewcommand{\theequation}{A.\arabic{equation}}
\setcounter{equation}{0}
 
 \renewcommand{\thesubsection}{A.\arabic{subsection}}
\setcounter{subsection}{0}

 \begin{figure}[h]
\epsfxsize=6cm
%\centerline{\epsfbox{nsigma22.eps}}
\centerline{\epsfbox{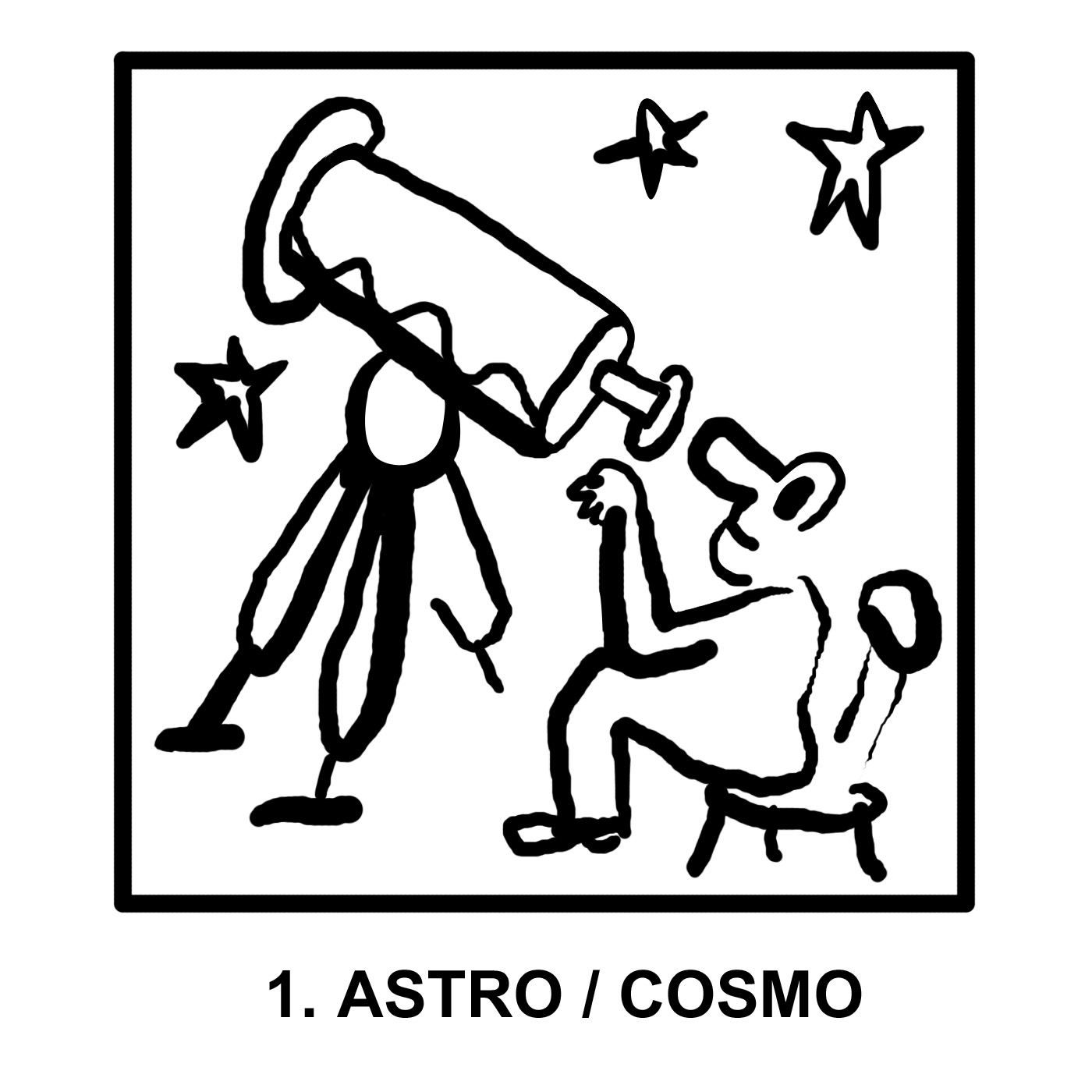}}
\caption{\small Astroparticle physics and cosmology. }
\label{astro}
\end{figure}
 \begin{figure}[h]
\epsfxsize=6cm
%\centerline{\epsfbox{nsigma22.eps}}
\centerline{\epsfbox{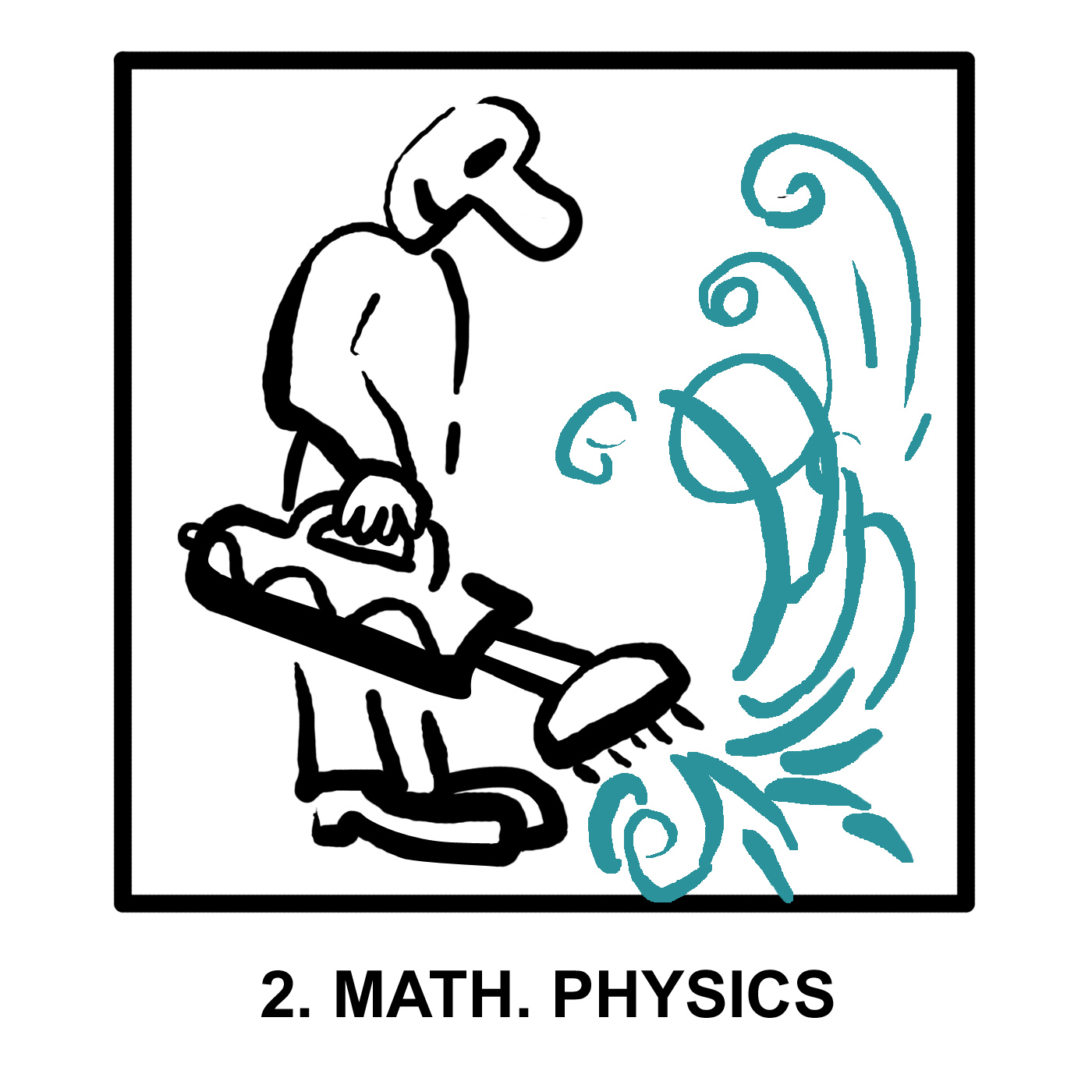}}
\caption{\small Mathematical physics absorbing string theory. }
\label{math}
\end{figure}
 \begin{figure}[h]
\epsfxsize=6cm
%\centerline{\epsfbox{nsigma22.eps}}
\centerline{\epsfbox{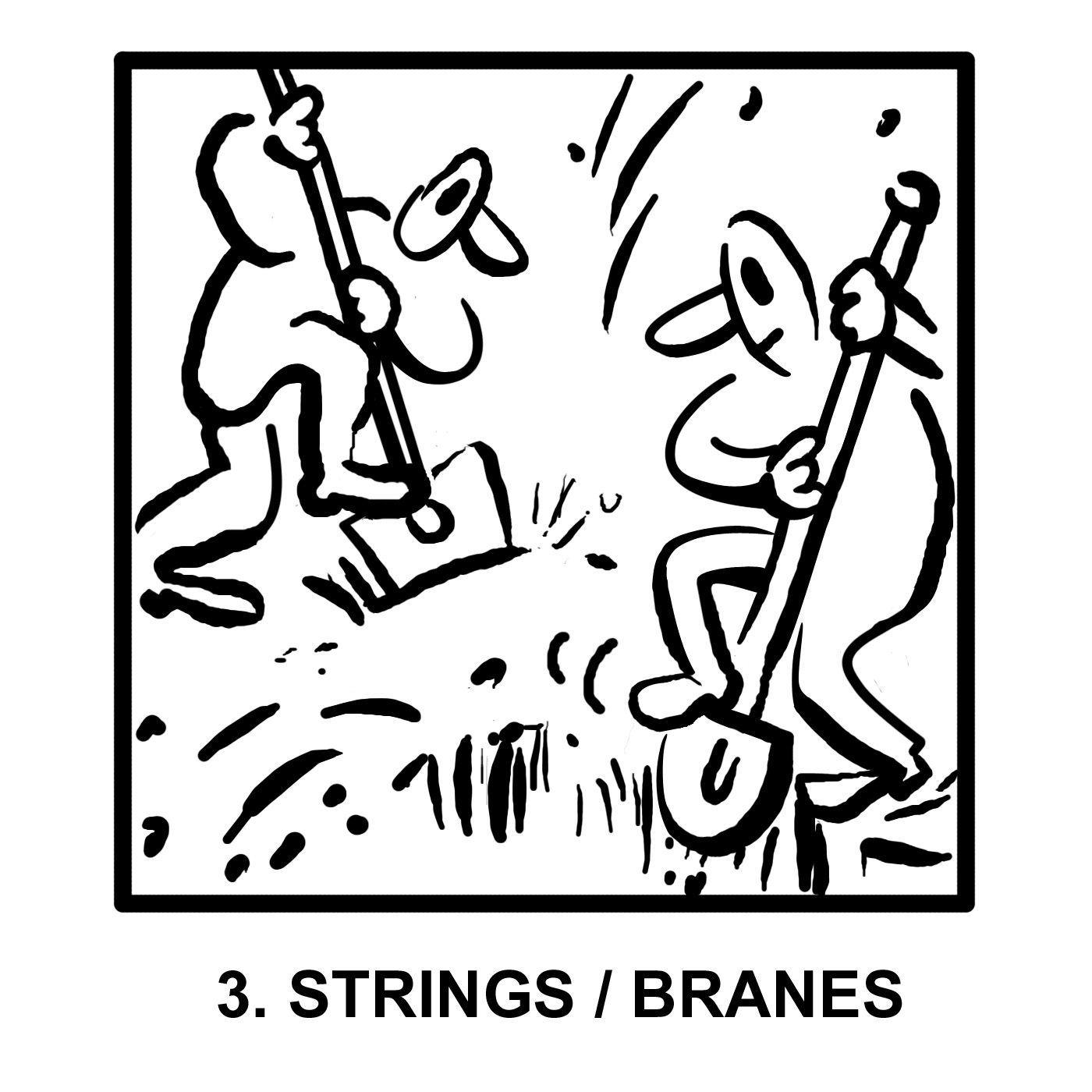}}
\caption{\small The theory of strings and branes. }
\label{strings}
\end{figure}
\begin{figure}[h]
\epsfxsize=6cm
%\centerline{\epsfbox{nsigma22.eps}}
\centerline{\epsfbox{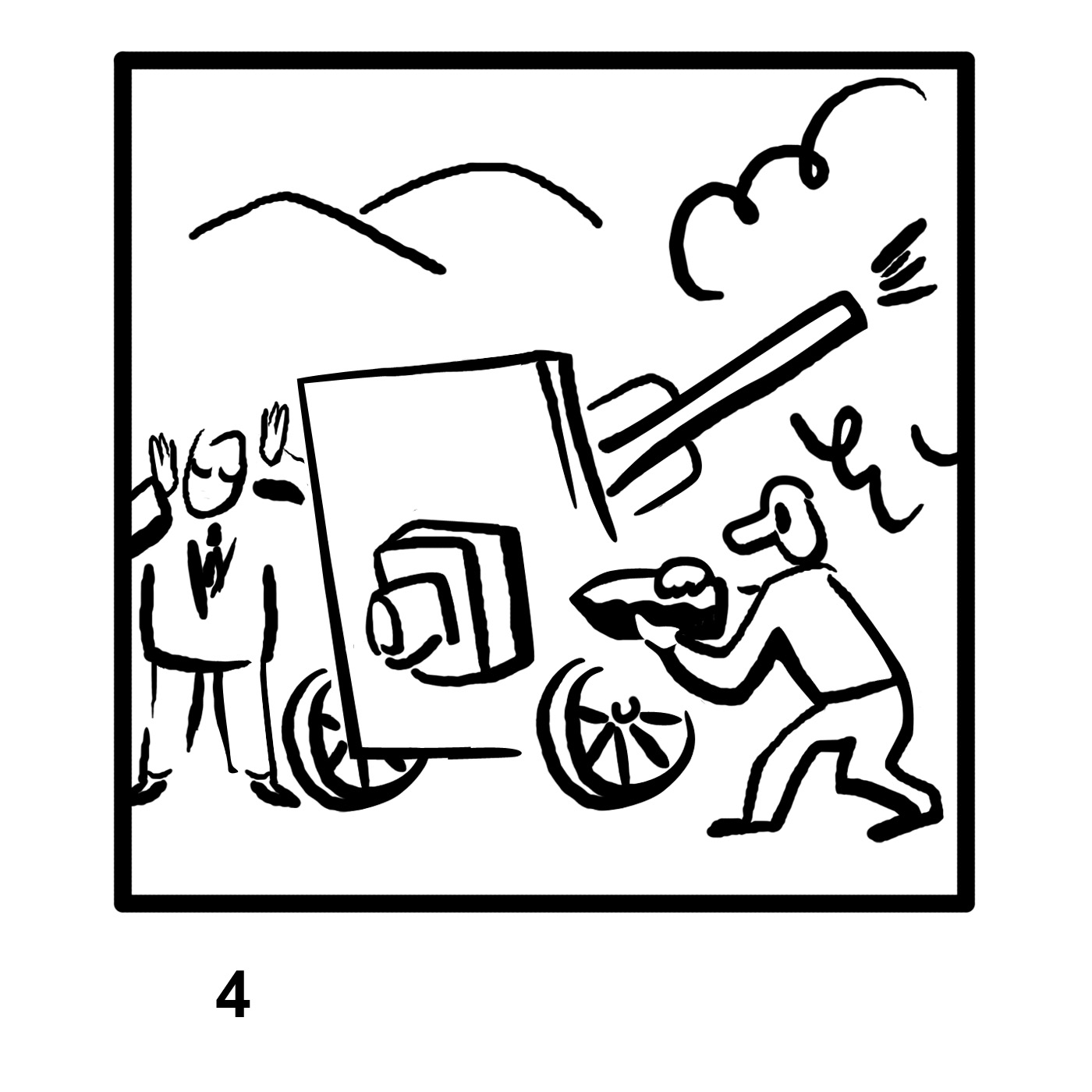}}
\caption{\small Supersymmetry-based phenomenology. }
\label{supheno}
\end{figure}
\begin{figure}[h]
\epsfxsize=6cm
%\centerline{\epsfbox{nsigma22.eps}}
\centerline{\epsfbox{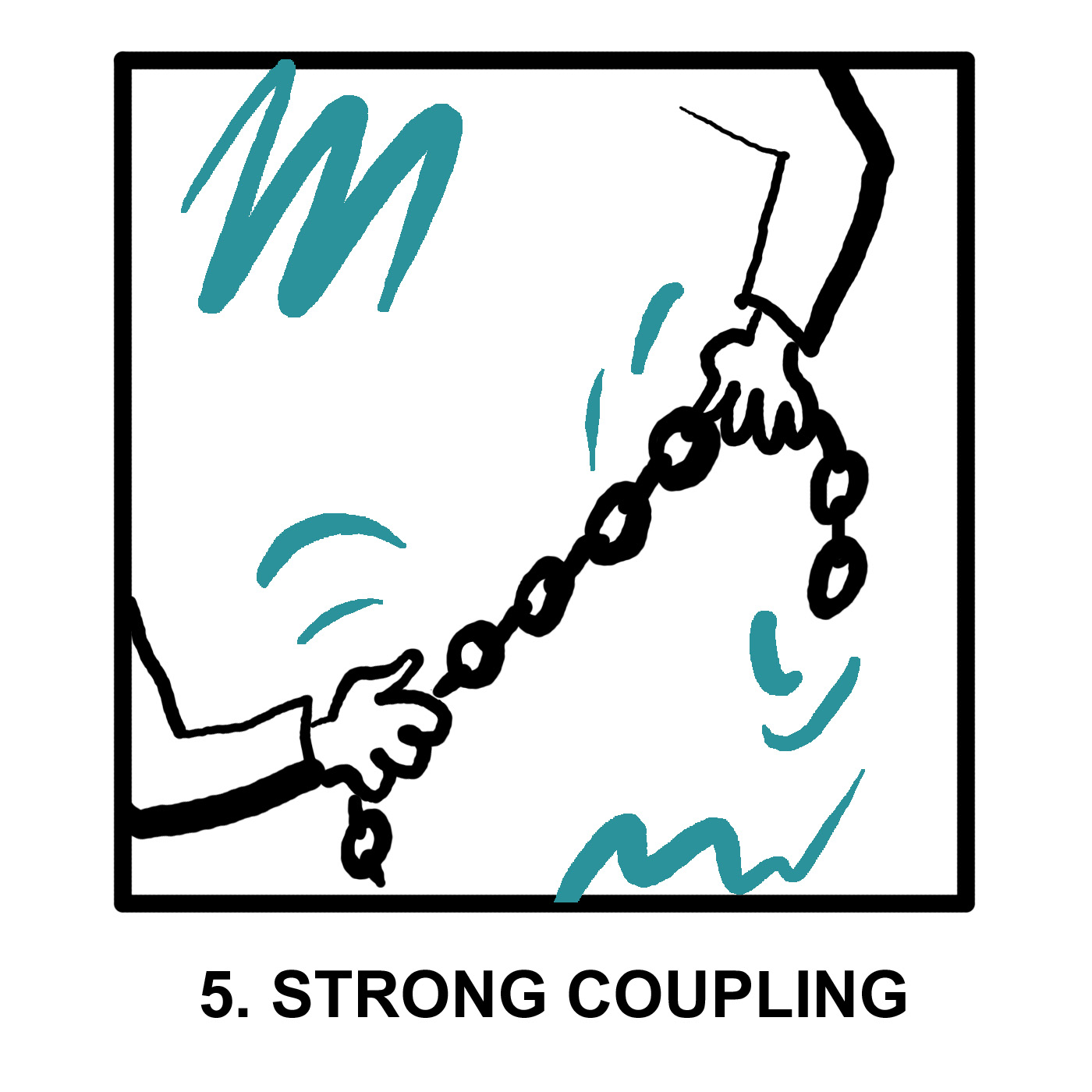}}
\caption{\small QCD and other gauge theories at strong coupling:
from confinement to chiral symmetry breaking and back via supersymmetry.}
\label{strong}
\end{figure}
\begin{figure}[h]
\epsfxsize=6cm
%\centerline{\epsfbox{nsigma22.eps}}
\centerline{\epsfbox{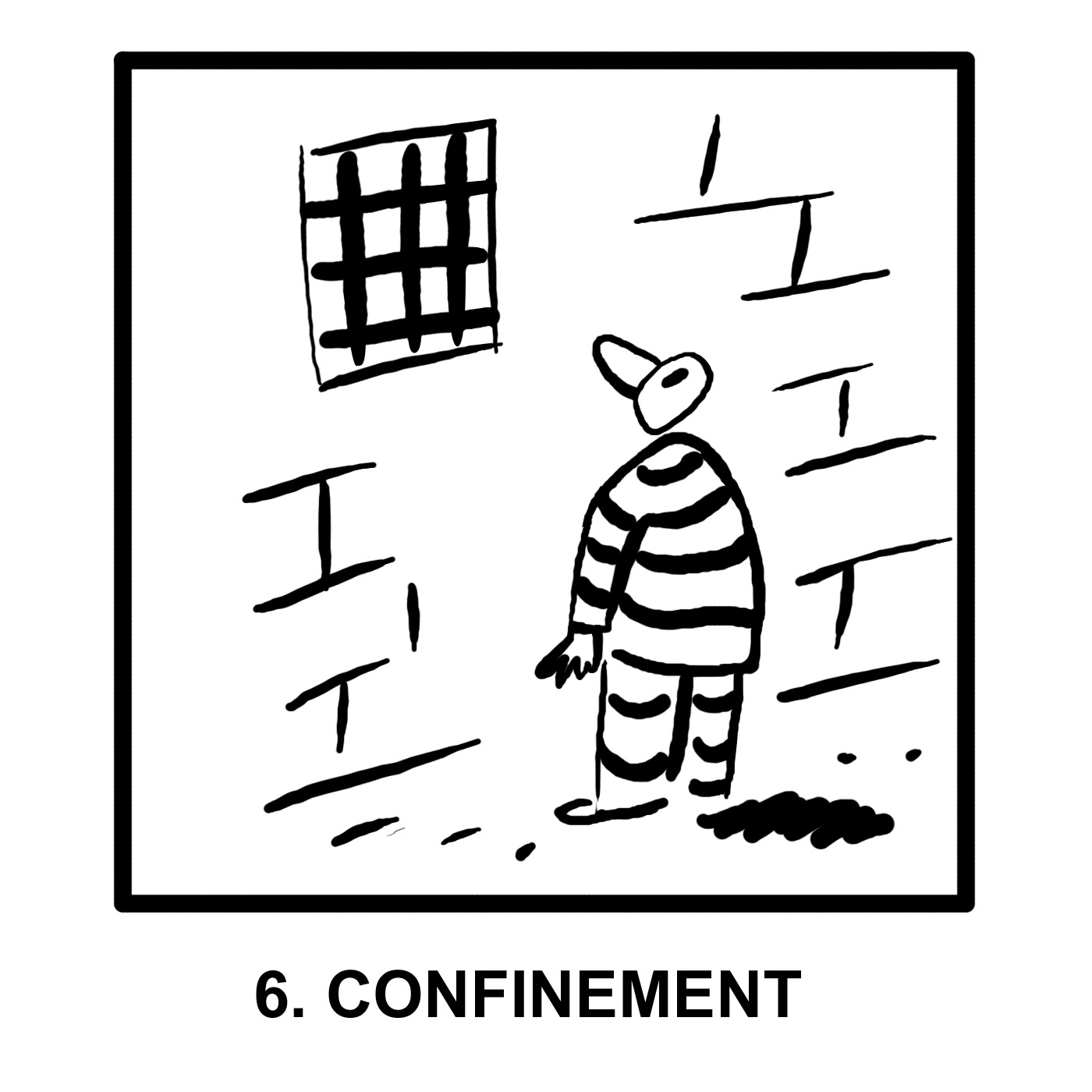}}
\caption{\small Deciphering mechanisms of confinement}
\label{confi}
\end{figure}
\begin{figure}[h]
\epsfxsize=6cm
%\centerline{\epsfbox{nsigma22.eps}}
\centerline{\epsfbox{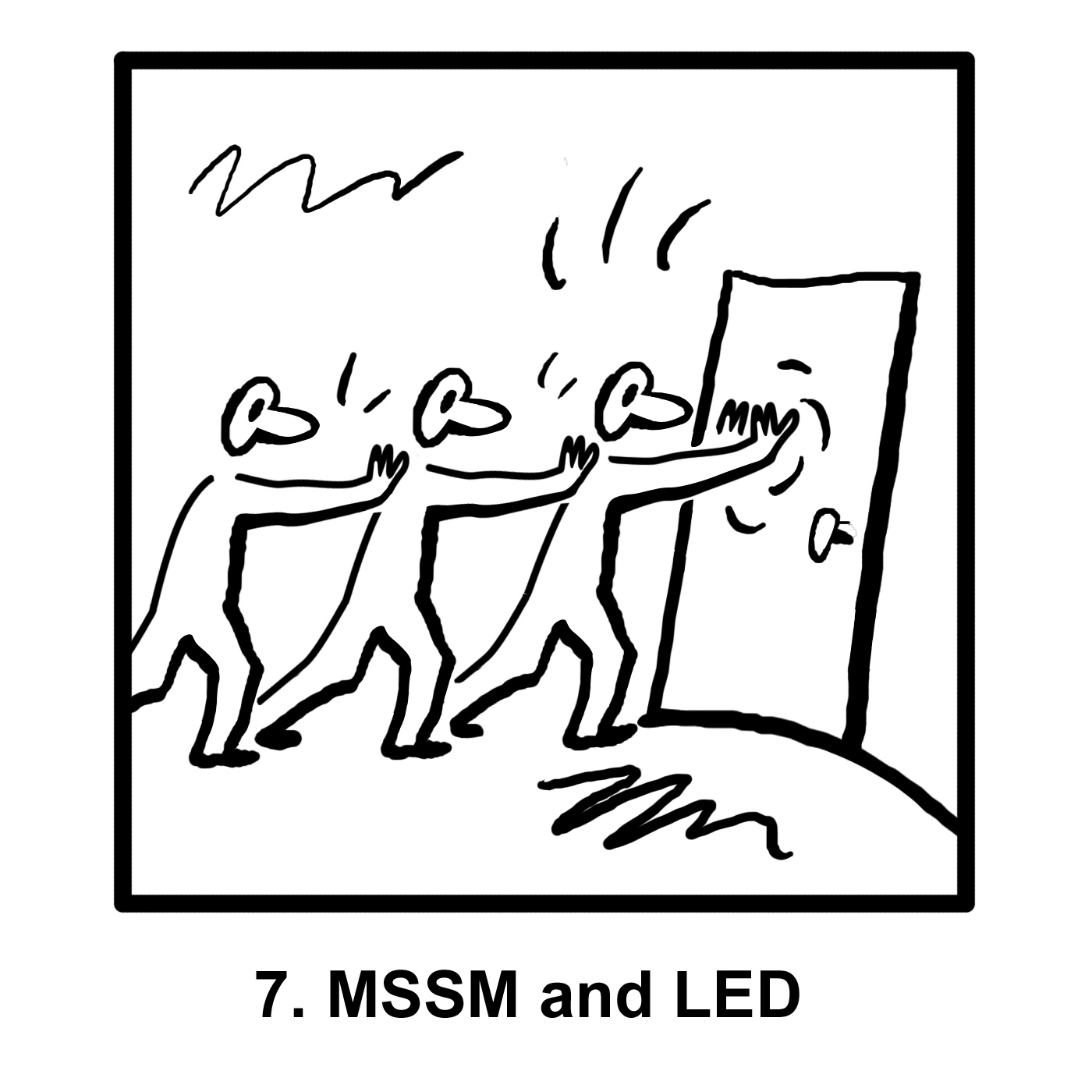}}
\caption{\small Searching for supersymmetry and/or large extra dimensions
``under a lamp post."}
\label{mssm}
\end{figure}
\begin{figure}[h]
\epsfxsize=6cm
%\centerline{\epsfbox{nsigma22.eps}}
\centerline{\epsfbox{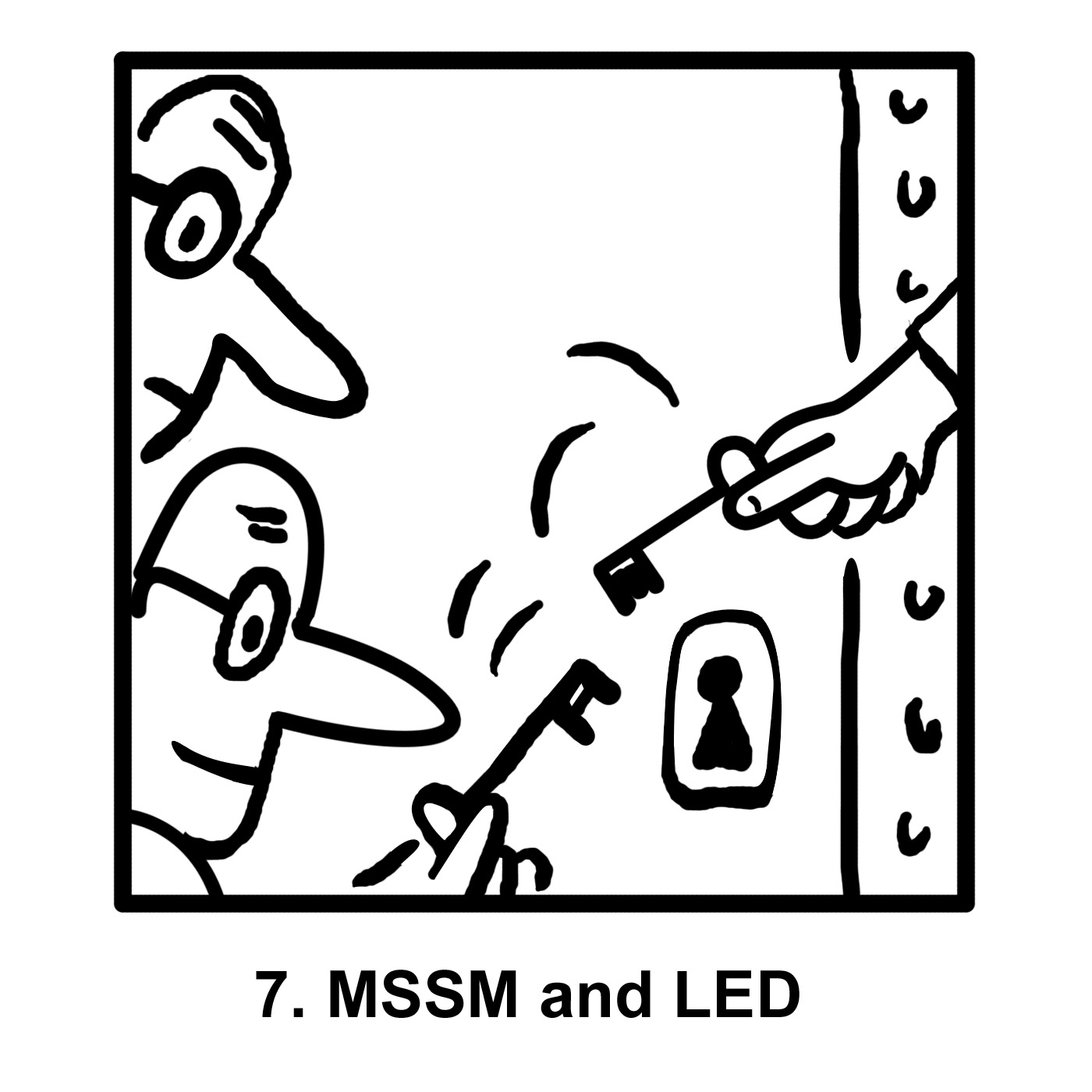}}
\caption{\small The search is continued in a different way.}
\label{moremssm}
\end{figure}
\begin{figure}[h]
\epsfxsize=6cm
%\centerline{\epsfbox{nsigma22.eps}}
\centerline{\epsfbox{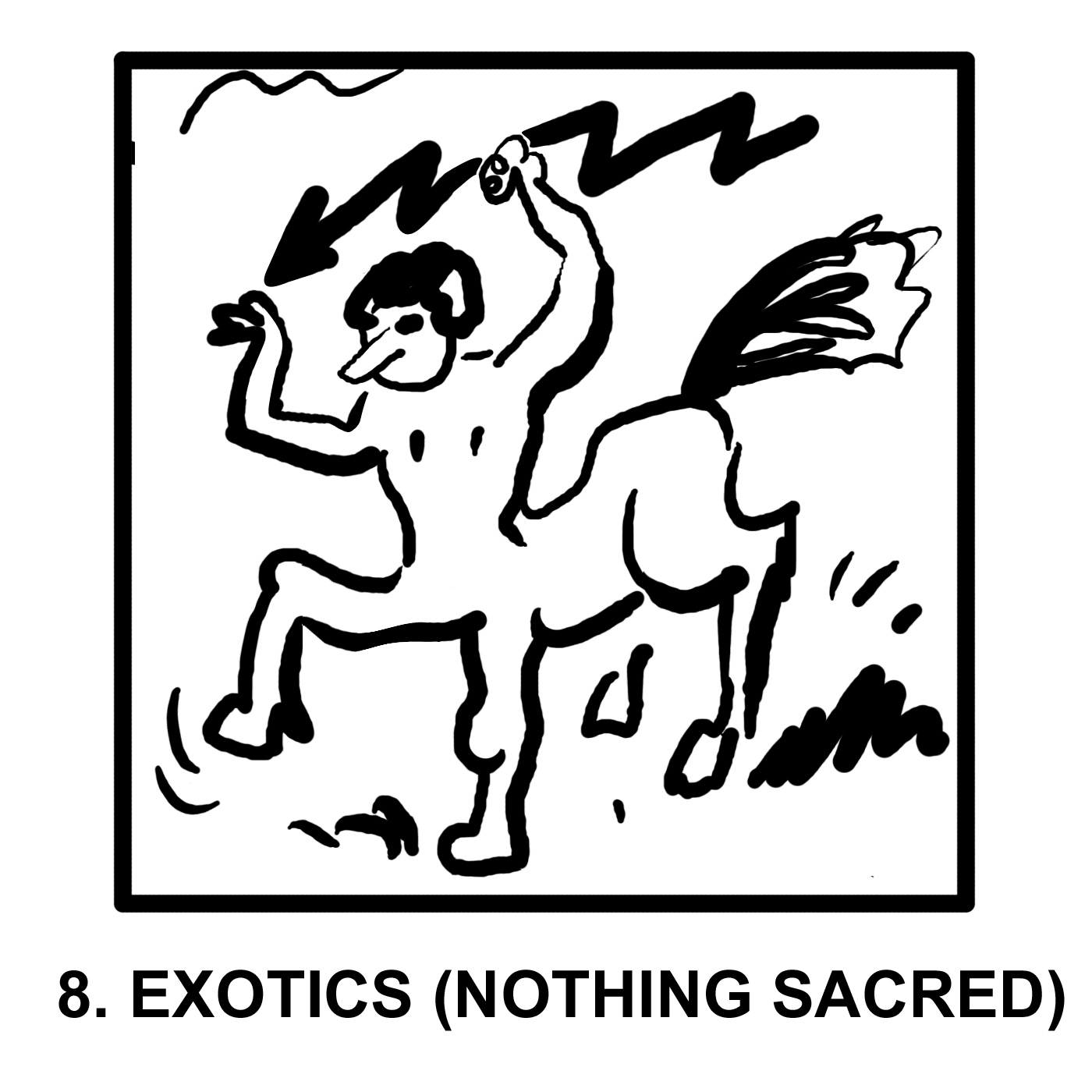}}
\caption{\small All things ``beyond" are highly appreciated. If we only new what exactly to look for ...}
\label{exo}
\end{figure}
\begin{figure}[h]
\epsfxsize=6cm
%\centerline{\epsfbox{nsigma22.eps}}
\centerline{\epsfbox{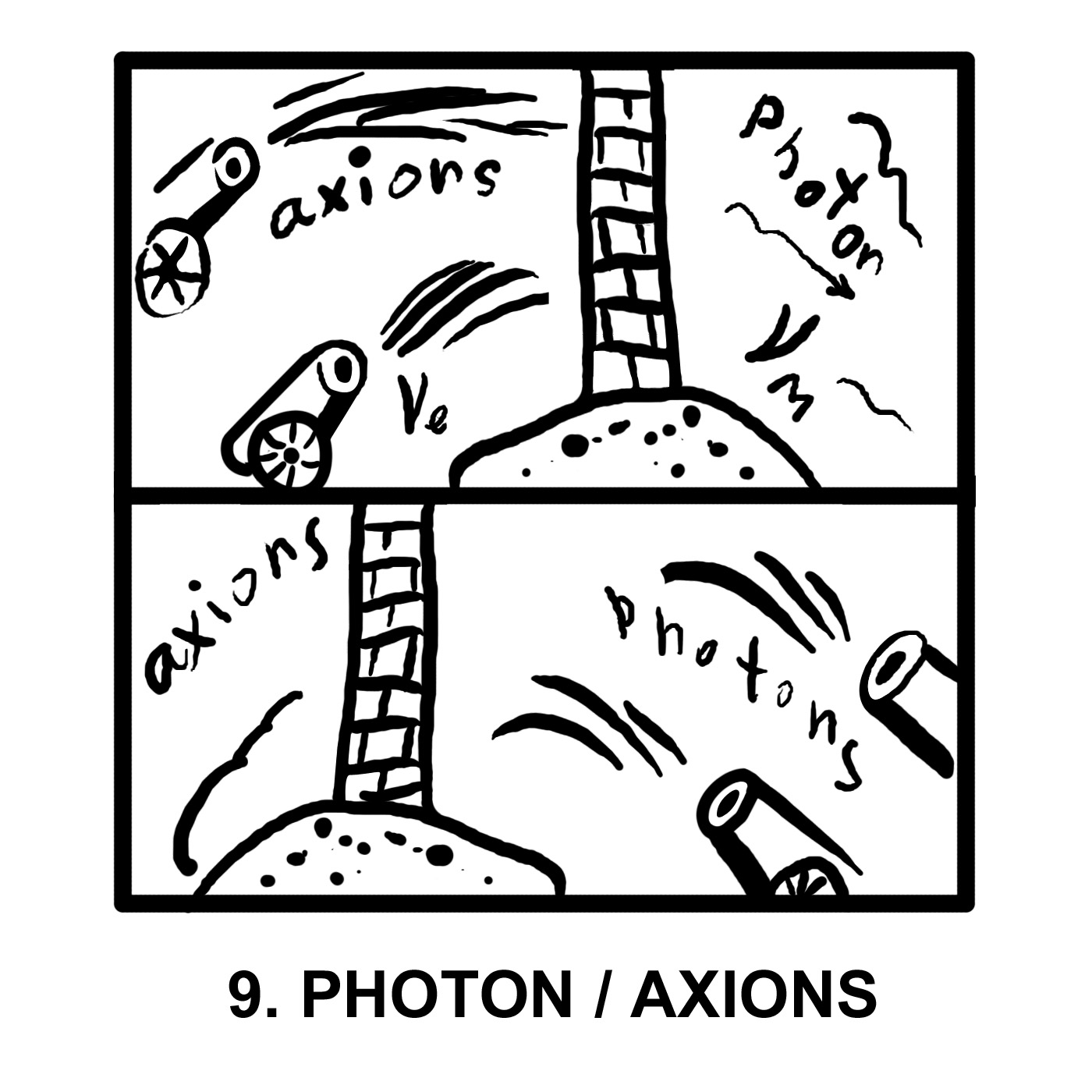}}
\caption{\small  Axion physics}
\label{axi}
\end{figure}
\begin{figure}[h]
\epsfxsize=6cm
%\centerline{\epsfbox{nsigma22.eps}}
\centerline{\epsfbox{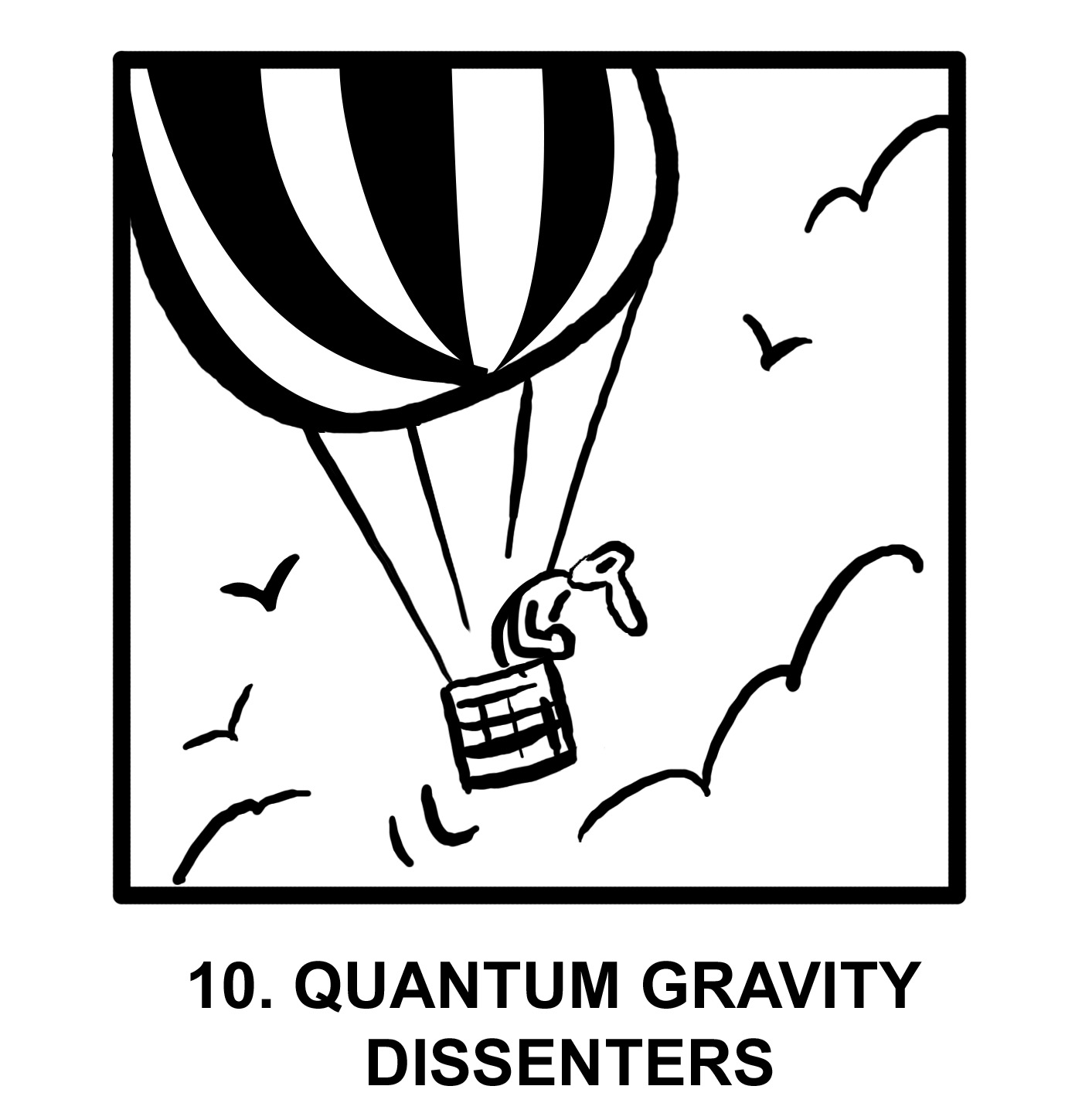}}
\caption{\small  The ever-lasting passion for quantum gravity}
\label{qgr}
\end{figure}

\end{document}